\documentclass[conference]{IEEEtran}
\renewcommand{\normalsize}{\fontsize{9.5}{11.5}\selectfont}
\usepackage{epsfig,rotating,setspace,latexsym,amsmath,epsf,amssymb,amsfonts,bm,theorem,epstopdf,cite,mathtools,caption,subcaption,enumerate,longtable,accents,bbm,algorithm,algorithmic,graphicx,epsf,authblk,url,color,multirow,comment,tabularx,dsfont,physics,tikz,dsfont}
\usepackage[inline]{enumitem}

\newtheorem{definition}{Definition}

\newtheorem{remark}{Remark}

\newcommand{\figref}[1]{\figurename~\ref{#1}}
\IEEEoverridecommandlockouts
\allowdisplaybreaks

\allowdisplaybreaks

\title{Temporary Power Adjusting Withholding Attack}

\author{Mustafa Doger \qquad Sennur Ulukus\\
\normalsize Department of Electrical and Computer Engineering\\
\normalsize University of Maryland, College Park, MD 20742\\
\normalsize  \emph{doger@umd.edu} \qquad \emph{ulukus@umd.edu}}

\date{}

\begin{document}

\maketitle
\begin{abstract}
We consider the block withholding attacks on pools, more specifically the state-of-the-art Power Adjusting Withholding (PAW) attack. We propose a generalization called Temporary PAW (T-PAW) where the adversary withholds a fPoW from pool mining at most $T$-time even when no other block is mined. We show that PAW attack corresponds to $T\to\infty$ and is not optimal. In fact, the extra reward of T-PAW compared to PAW improves by an unbounded factor as adversarial hash fraction $\alpha$, pool size $\beta$ and adversarial network influence $\gamma$ decreases. For example, the extra reward of T-PAW is 22 times that of PAW when an adversary targets a pool with $(\alpha,\beta,\gamma)=(0.05,0.05,0)$. We show that honest mining is sub-optimal to T-PAW even when there is no difficulty adjustment and the adversarial revenue increase is non-trivial, e.g., for most $(\alpha,\beta)$ at least $1\%$ within $2$ weeks in Bitcoin even when $\gamma=0$ (for PAW it was at most $0.01\%$). Hence, T-PAW exposes a significant structural weakness in pooled mining—its primary participants, small miners, are not only contributors but can easily turn into potential adversaries with immediate non-trivial benefits. 
\end{abstract}

\section{Introduction}
Nakamoto's Bitcoin (BTC) protocol, or more generally Nakamoto consensus, allows users to reach consensus in distributed permissionless settings \cite{btc-whitepaper}. Nakamoto's protocol incentivizes users by rewarding them for participating in a procedure called mining that is essentially solving a cryptographic puzzle. The (honest) protocol prescribes users to mine on top of the longest chain and share newly mined blocks immediately with their peers. When everyone follows the honest protocol, miners get rewards that is proportional to their fraction of hashpower in the system. A miner  working in such a blockchain with $y$ fraction of hashpower where a block is issued every $b$ minutes gets rewarded in $\frac{b}{y}$ minutes on average with variance $\frac{b^2(1-y)}{y^2}$. Hence, small miners join mining pools that regularly share the block rewards between the pool members to get a steady income rather than a fluctuating one with high variance \cite{rosenfeld2011analysisbitcoinpooledmining}.

Eyal and Sirer \cite{selfish-mining} showed that an entity can increase its share of rewards by adopting a mining strategy called \textit{selfish mining}. A myriad of works in the following decade delved deeper into this strategy from various points of view such as optimization, revenue changes and profit lags \cite{optimal-selfish,stubborn-mining,intermittent_mining,fork_after_witholding_attack,power_adjusting,doger2025incentiveattacksbtcshortterm,profit_lag,Grunspan_witholding_resilience,grunspan2019-profitability-selfish-mining}. Similar to selfish mining, another adversarial mining strategy, called \textit{Block Withholding} (BWH) attack, targeted at pool mining, increases an attacker's share of rewards \cite{courtois2014subversiveminerstrategiesblock}. In a nutshell, with a fraction of its mining power, an adversary joins a pool for mining but withholds the blocks it mines, but still gets rewards from the blocks mined by the other pool members. By carefully allocating its power between honest and targeted pool mining, an adversary can increase its fraction of the rewards \cite{power_splitting_pools}. In 2014, Eligius mining pool announced that it was subject to BWH attack and lost 300 BTC at the time \cite{EligiusBlockWithholding2014}.

An improved version of BWH is called \textit{Fork After Withholding} (FAW) attack where the adversary releases the withheld block when an honest miner (outside the target pool) mines a block to cause a fork race \cite{fork_after_witholding_attack}. \textit{Power Adjusting Withholding} (PAW) attack is a further improvement, where the adversary readjusts its power allocation to pool mining after it finds a block that increases the extra reward (up to 2.5 times the extra reward in FAW) \cite{power_adjusting}. An in-depth analysis of PAW is given in \cite{doger2025incentiveattacksbtcshortterm} which derives PAW-related quantities in an explicit manner without the simplifying assumptions of \cite{power_adjusting} related to fork races. 

Authors of \cite{Grunspan_witholding_resilience} show that selfish mining is an attack on the difficulty adjustment algorithm (DAA) of BTC and a temporal revenue analysis shows that selfish mining strategy is sub-optimal compared to honest mining absent any DAA. The analysis of \cite{doger2025incentiveattacksbtcshortterm} discovers that even though BWH and FAW attacks are also sub-optimal, PAW strategy dominates the honest mining strategy for some parameters. Further, the authors prove that the honest miners outside the target pool profit from the PAW attack, in cases even more than the adversary mounting the attack. 

In this paper, we propose \textit{Temporary PAW} (T-PAW) which is essentially the same strategy as PAW however there is a deadline on block withholding, i.e., the adversary releases the withheld block after at most $T$-time. Our results are as follows:
\begin{itemize}
    \item We show that the PAW strategies are sub-optimal to T-PAW and there is a large space of improvement available for small miners. In fact, the extra reward of T-PAW compared to PAW increases by an unbounded factor as adversarial hash fraction $\alpha$, pool size $\beta$ and adversarial network influence $\gamma$ decreases.
    \item We replace an approximation used in previous PAW analysis \cite{power_adjusting,doger2025incentiveattacksbtcshortterm} regarding the reward sharing in pools with an exact analysis using stochastic distributions of mining times.
    \item T-PAW attack results in no short-term profit lag for the adversary in most parameter regimes even when the goal is to maximize its revenue ratio. 
    \item T-PAW attack's profitability does not require a DAA and dominates honest mining. The adversary can get non-trivial revenue increase for all parameters even when $\gamma=0$. 
    \item PAW attack is shown to benefit the honest miners outside the target pool, in many cases even more than the adversary launching the attack \cite{doger2025incentiveattacksbtcshortterm}. In T-PAW, even when $\gamma=0$, the adversary benefits most from the attack.
\end{itemize}
The results above suggest that unlike PAW (and previous BWH attack types), even a small miner can get non-trivial benefits from T-PAW. When one also considers the fact that the pools are formed primarily by small home miners, the attack exposes a structural weakness in pooled mining—its primary participants, small miners, are not only contributors but can easily turn into potential adversaries with non-trivial and immediate gains.

\section{System Model}
We consider a blockchain functioning under the Nakamoto longest chain Proof-of-Work (PoW) protocol with $n$ miners where $n$ is large, i.e., each miner controls an infinitesimal fraction of hashpower in the system. We consider $\alpha$ fraction of the miners to be adversarial and colluding, which in turn can be represented as a single entity, called \textit{the adversary}, controlling $\alpha$ fraction of the hashpower. The remaining miners are assumed to be honest, i.e., they mine on top of the longest chain and immediately share the newly mined blocks with the network. We assume that if everyone including the adversary mines honestly, the block arrival process follows the exponential distribution with $\mathrm{Exp}(\lambda_1)$.

We are interested in BWH against pools, hence we assume a target pool that consists of honest miners and makes up $\beta$ fraction of the total hashpower in the system. The pool is run by a manager who assigns PoW tasks to its miners, which are analogous to puzzles. The miners working towards their assignments encounter fPoW (full PoW) and pPoW (partial PoW) solutions that they regularly share with the manager. pPoW solutions, which can be seen as easier versions of the puzzle assignment, help the manager determine the contribution of each member towards the mining. On the other hand, fPoW solutions essentially create valid blocks and the manager shares the reward between the miners according to their contribution. We assume that the adversary attacks the target pool of size $\beta$ under the condition that $\alpha+\beta<0.5$. More specifically, the adversary allocates some fraction of its hashpower to join the target pool and mounts a sophisticated BWH attack where it shares pPoW regularly but withholds fPoW according to some predetermined strategy to be explained in Section~\ref{sec::t-paw}.

In this paper, following the convention in the literature \cite{selfish-mining,intermittent_mining,optimal-selfish,courtois2014subversiveminerstrategiesblock,fork_after_witholding_attack,power_adjusting,grunspan2019-profitability-selfish-mining,doger2025incentiveattacksbtcshortterm,profit_lag,power_splitting_pools,stubborn-mining,Grunspan_witholding_resilience}, we assume that the adversary becomes aware of any block as soon as it is mined and we ignore the innocent forks, i.e., the forks between the \textit{honest blocks} (mined by honest miners) caused by network delays. On the other hand, forks caused by adversarial strategies are treated as follows. Assume the adversary withholds a fPoW mined as part of pool mining and later an honest block is created and shared by an honest miner outside the pool. In such a case, the adversary can rush to release the withheld fPoW and $\gamma$ fraction of the honest miners outside the pool receive the adversarial fPoW first and mine on top of it, whereas $1-\gamma$ fraction of the honest miners outside the pool receive the honest block first and mine on top of it. All the honest miners of the target pool follow the decision of the pool manager. If the pool manager is rational ($\mathds{1}_{P_R}=1$), it picks the adversarial fPoW, which increases the revenue of the pool compared to the case of $\mathds{1}_{P_R}=0$.

Further, we assume that the blockchain adopts a DAA similar to that of BTC. Let us call the blocks that become part of the longest chain in the long run as \textit{canonical blocks}. The blockchain readjusts the mining difficulty after every $D_0$ canonical blocks are mined to maintain a steady block production rate of $\frac{D_0}{\tau_0}$ where the times between adjustments are called \textit{epochs}. More specifically, given the current mining difficulty of $d_c$ in an epoch and the epoch lasts $\tau_c$ time to produce $D_0$ blocks, the blockchain readjusts the difficulty at the beginning of the next epoch as $d_n=d_c\frac{\tau_0}{\tau_c}$.\footnote{For the sake of convenience, we assume that a higher difficulty means fewer blocks mined. Further, even though many DAAs such as that of BTC restrict the change to be between $d_n\in[1/4d_c,4d_c]$, our system model guarantees $d_n\in[1/2d_c,2d_c]$.} 

\begin{table}[t]    
    \begin{center}
\begin{tabular}{||c | c||} 
 \hline
 Parameters & Values \\ [0.5ex] 
 \hline\hline
 $\alpha$ & adversarial hash fraction  \\ 
 \hline
 $\beta$ & target pool hash fraction  \\ 
  \hline
 $p_i$ ($i=1,2$) & fraction of $\alpha$ at pool mining\\
 \hline
 $\lambda_1$ & total block generation rate \\
  \hline 
 $\lambda_2$ & $(1-\alpha p_2)\lambda_1$  \\
 \hline
 $a_i$ ($i=1,2$) & $\beta+\alpha p_i$  \\ 
 \hline 
 $a_{i}'$ ($i=1,2$) & $\frac{\alpha p_i}{a_i}$ \\
 \hline
 $\lambda'_i$ ($i=1,2$) & $\frac{\lambda_i}{a_i}$\\ 
 \hline
  $T'$ & $a_2T$\\
 \hline
  $E_1(t)$ & $\int_t^{\infty}\frac{e^{-x}}{x}dx$\\
 \hline
\end{tabular}
\end{center}
\caption{ Frequently used notations.}
\label{table::freq-table}\vspace{-.5cm}
\end{table}

\subsection{Definitions}
Here, we list a series of notations used frequently throughout the paper in Table~\ref{table::freq-table} and define the following useful metrics.  
\begin{definition}
    \textbf{Relative Extra Rewards (RER)} \cite{power_adjusting} of an entity $x$ ($RER_x^{S_1,S_2}$) is defined as
\begin{align}
    RER_x^{S_1,S_2}=\frac{\rho_x^{S_1}-\rho_x^{S_2}}{\rho_x^{S_2}}
\end{align}
where $\rho_x^{S}$ represents the revenue ratio of an entity $x$ when the adversary adopts the strategy $S$. The \textbf{Revenue Ratio} of an entity is defined as the fraction of the rewards the entity receives in the long-run out of the total rewards issued in the system.
\end{definition}

\begin{definition}
    \textbf{Revenue Change} \cite{doger2025incentiveattacksbtcshortterm,profit_lag} of an entity $x$ at time $t$ when the adversary adopts the strategy $S$ compared to when the adversary adopts the honest strategy $H$ is
\begin{align}
   \Delta_x^{S,H}(t)=\begin{cases} 
  (\rho_x^{S}-\rho_x^{H}\delta^{S})\frac{t}{\delta^S\tau_0}, & t\leq \delta^S\tau_0 \\ 
  \rho_x^{S}-\rho_x^{H}\delta^{S}+(\rho_x^{S}-\rho_x^{H})\frac{t-\delta^S\tau_0}{\tau_0}, & t>\delta^S\tau_0 
\end{cases}
\end{align}
where $\rho_x^{H}$ is the fraction of hashpower of entity $x$ in the system whereas $\delta^{S}$ represents the block redundancy ratio when the adversary adopts the strategy $S$.\footnote{Throughout the paper, we consider a reference time $t_0=0$ as the starting time of a new epoch and the adversary mounts a strategy $S$ at $t_0$, before which everyone including the adversary was following the honest protocol. For the notion of revenue change, we assume that the total rewards issued in an epoch is scaled to $1$ unit, i.e., 1 canonical block rewards $1/D_0$ rewards.}
Similarly, 
\textbf{Relative Revenue Change} is defined as
    $\overline\Delta_x^{S,H}(t)=\frac{\Delta_x^{S,H}(t)}{\rho_x^{H}}$.
\end{definition}

\begin{definition}
    \textbf{Profit Lag} \cite{doger2025incentiveattacksbtcshortterm,profit_lag} of an entity $x$ when the adversary adopts the strategy $S$ is defined as
\begin{align}
    \Delta_x^{S,H}=\inf \{\tau:\Delta_x^{S,H}(t)>0, \forall t>\tau\}.
\end{align}
\end{definition}

\begin{remark}
The revenue change of the adversary $A$ at $t_1=\delta^S\tau_0$ tells us how much its revenue changes at the end of the first epoch compared to the case if it adopted the honest strategy $H$. It should be clear that $\Delta_x^{H,H}(t)=0$ for $\forall (t,x)$ since $\delta^H=1$. If $\Delta_A^{S,H}(t_1)>0$, the profit lag of the adversary is zero, i.e., under strategy $S$ the adversary gets more revenue than honest mining even when there is no difficulty adjustments in the system. We refer the reader to \cite{power_adjusting,doger2025incentiveattacksbtcshortterm,profit_lag} for a more detailed explanation to understand the intuition behind the metrics defined above. 
\end{remark}

\section{Temporary Power Adjusting Withholding}
Before introducing our attack explicitly, we briefly revisit the previous PAW models \cite{power_adjusting,doger2025incentiveattacksbtcshortterm}. The model and analysis of PAW used in \cite{power_adjusting} assumes some simplifications in the fork races pertaining to the attack whereas a more rigorous analysis of the attack is provided in \cite{doger2025incentiveattacksbtcshortterm}. The simplified model assumes that each attack cycle contributes one block to the longest chain growth and is independent from other cycles whereas the latter model considers the fork races explicitly until the attack cycles fully end. Since we build complicated versions of PAW attack that involve multiple block growth in each attack cycle, in this paper, we use the rigorous modeling of \cite{doger2025incentiveattacksbtcshortterm}. Further, although \cite{doger2025incentiveattacksbtcshortterm} builds a more rigorous analysis compared to \cite{power_adjusting}, both models still use an approximation regarding the rewards of the fPoWs shared between pool members and the adversary (see Appendix~\ref{sec::app::paw-reduction}). The approximation error grows large especially for T-PAW attacks considered in this paper. Hence, we provide an explicit and rigorous analysis regarding the rewards of the fPoWs, which enables us to derive exact results rather than approximations. 

\subsection{Attack Description: T-PAW}\label{sec::t-paw}
\begin{figure}[t]
\centering
\includegraphics[width=\columnwidth]{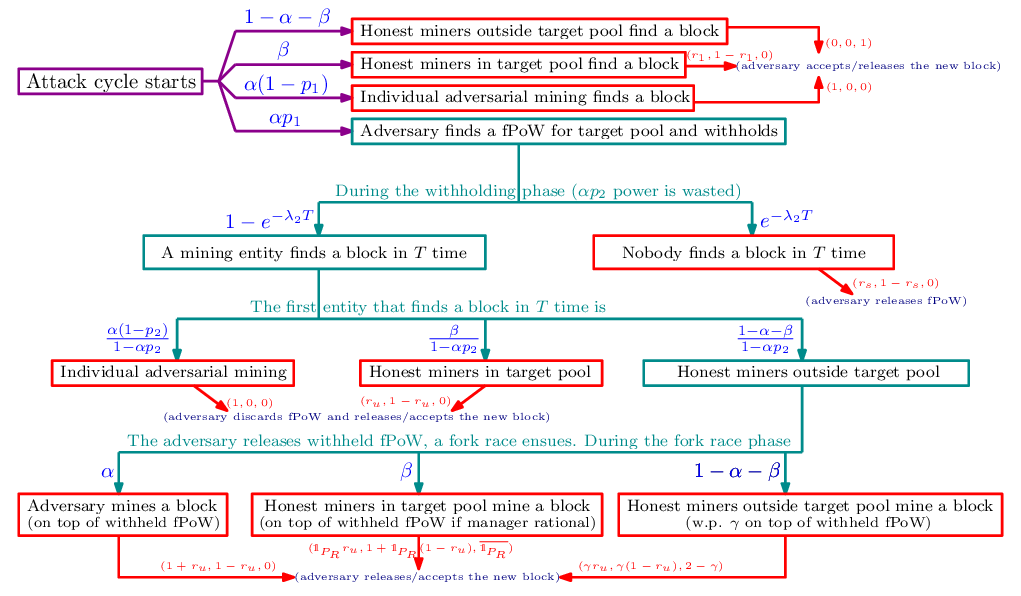}
    \caption{Evolution of a T-PAW attack cycle.}
    \label{fig::cycle_events}
\end{figure}

Let us start by explicitly stating the  T-PAW attack. Let $\lambda_1$ denote the total mining rate of the full network, i.e., when everyone mines honestly. At the start of the attack cycle, the adversary mines individually and honestly with $(1-p_1)$ fraction of its hashpower, i.e., mines on the tip of the longest chain and releases the block immediately. With remaining $p_1$ fraction, it mines as part of the pool. It regularly submits the pPoWs until it encounters a fPoW. \begin{itemize*}
    \item If another miner finds a block before the adversary encounters a fPoW, the adversary accepts the new block and starts a new attack cycle on top of the new block.
    \item If the adversary encounters a fPoW, it withholds the fPoW and readjusts $p_1$ to $p_2$. The ensuing phase of the attack is called the withholding phase, during which:
\end{itemize*}
\begin{enumerate}
    \item If the adversary finds another fPoW with $p_2$ fraction, it discards the old one and keeps holding the new one.\label{step::1-fPoW}
    \item If an honest pool miner finds a block, the adversary accepts the block and discards the withheld fPoW.\label{step::2-honest}
    \item If the adversary mines a block with its individual mining power, it releases this new block and discards the fPoW.\label{step::3-adv}
    \item If an honest miner outside the pool finds a new block, the adversary releases the withheld fPoW and a fork race ensues (fork race phase), where the adversary, the pool (if the pool manager is rational, i.e., $\mathds{1}_{P_R}=1$) and $\gamma$ fraction of the miners outside the pool prefer the adversarially released fPoW. During this fork race, the adversary mines honestly and individually on top of the fPoW with all its power.\label{step::4-fork}
    \item If none of the Cases~\ref{step::2-honest}, \ref{step::3-adv}, \ref{step::4-fork} happens in $T$ time (we do not care about Case~\ref{step::1-fPoW}), the adversary releases the fPoW. 
\end{enumerate}
After one of the Cases above (except Case~\ref{step::1-fPoW}) happens (and after the fork races resolve if Case~\ref{step::4-fork} happens), the attack cycle ends and the adversary reverts its power allocation to pool as $p_1$ and a new attack cycle starts. In \figref{fig::cycle_events}, we display all the paths an attack cycle can take. The cyan boxes represent middle steps within a cycle that ends in one of the red boxes where the quantities in blue represent the probabilities and in red represent the rewards of each entity (the adversary, honest pool miners, honest non-pool miners) at the end of the cycle (see Appendix~\ref{sec::app::t-paw} for derivations). 

\subsection{Relation to Existing Strategies}\label{sec::discussion_tpaw_paw}
First, note that when $T=0$, the T-PAW reduces to honest mining strategy irrespective of $p_1,p_2$ values. On the other hand, when $T\to\infty$, T-PAW reduces to PAW described in \cite{doger2025incentiveattacksbtcshortterm}.\footnote{The (simplified fork race) PAW modeling of \cite{power_adjusting} with $c=\gamma(1-\alpha-\beta)+\alpha+\beta\mathds{1}_{P_R}$ results in approximately the same revenue ratios as (rigorous) PAW modeling of \cite{doger2025incentiveattacksbtcshortterm}. PAW modeling of \cite{power_adjusting} reduces to FAW \cite{fork_after_witholding_attack} when $p_1=p_2$. FAW attack in turn reduces to BWH when $c$ is set to $0$  (see Appendix~\ref{sec::app::paw_simple}).} Our analysis suggests that for many parameters, especially when $\alpha,\beta,\gamma$ are small, $T\to\infty$ is not optimal and there is a large room for improvement compared to PAW. Such an argument might be counterintuitive when one takes the memorylessness property of exponential arrivals into account since the first arrival time in Cases~\ref{step::2-honest}, \ref{step::3-adv}, \ref{step::4-fork} does not depend on time $T$, i.e., how long the adversary already withheld the fPoW. However, the reward of the adversary from the fPoW is not memoryless, which effects the final revenue ratio. For example, withholding the fPoW and increasing $p_1$ to $p_2$ increases the adversarial fractional reward, but as more time passes the fractional reward saturates and the adversary only risks its share of rewards from the fPoW.

It is shown in \cite{doger2025incentiveattacksbtcshortterm} that for many parameters, the PAW attack can be superior to honest mining even when there is no DAA in the protocol. On the other hand, without DAA, withholding strategies such as selfish mining result in less revenue than honest mining \cite{Grunspan_witholding_resilience}. An inspection of such a situation in \cite[Example~1]{doger2025incentiveattacksbtcshortterm} with $T=\infty$ suggests that the phenomenon occurs especially when $p_2$ is large and $\beta$ is small and $\gamma$ is large. In such a situation, the adversary significantly increases $p_1$ to $p_2$, since if the fork race resolves in its favor, it will get more revenues from fPoW than it would if it were to act honestly and publish immediately.

Notice, when an attack cycle starts and the adversary encounters fPoW, it withholds and readjusts its power allocation to the pool until some other entity finds a block in the regular PAW attack ($T=\infty$). There are two shortcomings of this strategy even if $p_2$ is large and $\beta$ is small: \begin{enumerate*}
\item If it takes a long while until someone else finds a block, the total adversarial fractional contribution to the pool in that attack cycle is already saturated.  
\item If $\gamma$ is not large, the adversary is risking all the contributions to fPoW by waiting more since the fork race is less likely to result in its favor.
\end{enumerate*} 
Thus, we propose that the adversary withholds the fPoW at most for a finite time $T$ that is to be optimized in addition to $p_1$, $p_2$, which in turn brings improvements.

\section{Analysis of T-PAW Quantities}
Let us denote the adversarial revenue ratio as $\rho^{p_1,p_2,T}_{A}$ (or simply $\rho_A$ as we drop the superscripts notation for convenience). Let $B_A$ denote the reward the adversary gets in an attack cycle whereas $B_C$ denote the total reward in an attack cycle from the canonical blocks. Since each attack cycle is independent, due to the law of large numbers, we have, 
\begin{align}
    \rho_A=\rho^{p_1,p_2,T}_{A}=&\frac{\mathbb{E}[B_A]}{\mathbb{E}[B_C]},\label{eq::adv_rho_bwh_type_b_t}
\end{align}
where
\begin{align}
\mathbb{E}&[B_C]=1+\alpha p_1\Big(1-e^{-\lambda_2 T}\Big)\frac{1-\alpha-\beta}{1 - \alpha p_2},\label{eq::canonical_bwh_type_b_t}\\
\mathbb{E}&[B_A] =  \alpha(1-p_1) + \beta r_1 + \alpha p_1\bigg(e^{-\lambda_2 T}r_s \nonumber \\
&+\big(1-e^{-\lambda_2 T}\big)\Big( \frac{\alpha(1 - p_2)}{1 - \alpha p_2} + \frac{\beta}{1 - \alpha p_2}r_u  \nonumber \\
&+\frac{1-\alpha-\beta}{1 - \alpha p_2}\big(r_u(\gamma(1-\alpha-\beta)+\alpha+\beta\mathds{1}_{P_R})+\alpha\big)\Big)\bigg),\label{eq::adv_rho_bwh_type_b_t_Ba}
\end{align}
with $r_1=a_1'$,
\begin{align}
r_s&=a_1'+(a_2'-a_1')\lambda'_1 T' e^{\lambda'_1 T'}E_{1}(\lambda'_1 T'),\\
r_u&=a_1'+(a_2'-a_1')\frac{\lambda'_1 \lambda'_2}{1 - e^{-\lambda'_2 T'}} 
\int_0^{T'} t \, e^{(\lambda'_1 - \lambda'_2)t} E_1(\lambda'_1 t)\, dt.
\end{align}
Similarly, it can be shown that the revenue ratio of the rest of the pool members is
\begin{align}
    \rho_{pool}=\rho^{p_1,p_2,T}_{pool}=\frac{\mathbb{E}[B_P]}{\mathbb{E}[B_C]},
\end{align}
where $B_P$ denotes the reward the honest pool miners get in an attack cycle from the canonical blocks and its expectation is
\begin{align}
\mathbb{E}&[B_P] = \beta (1-r_1) + \alpha p_1\Bigg(e^{-\lambda_2 T}(1-r_s) \nonumber \\
&+\big(1-e^{-\lambda_2 T}\big)\Big( \frac{\beta}{1 - \alpha p_2}(1-r_u) +\frac{1-\alpha-\beta}{1 - \alpha p_2} \nonumber \\
&\times\big((1-r_u)(\gamma(1-\alpha-\beta)+\alpha+\beta\mathds{1}_{P_R})+\beta\big)\Big)\Bigg).\label{eq::adv_rho_bwh_type_b_t_Bp}
\end{align}
For the honest miners outside the pool, the revenue ratio is
\begin{align}
    \rho_{rest}=\rho^{p_1,p_2,T}_{rest}=\frac{\mathbb{E}[B_R]}{\mathbb{E}[B_C]},\label{eq::rho_rest_type_b_t}
\end{align}
where
\begin{align}
\mathbb{E}[B_R] =& (1-\alpha-\beta)\bigg(1+\big(1-e^{-\lambda_2 T}\big)\frac{\alpha p_1}{1-\alpha p_2}\nonumber\\
&\times\big((1-\alpha-\beta)(2-\gamma)+\beta \overline{\mathds{1}_{P_R}}\big)\bigg),\label{eq::adv_rho_bwh_type_b_t_Br}
\end{align}
and $\overline{\mathds{1}_{P_R}}=1-\mathds{1}_{P_R}$. Note that, $\rho_{rest}\geq1-\alpha-\beta$ independent of the values $p_1,p_2,T$ since $\mathbb{E}[B_R]\geq (1-\alpha-\beta)\mathbb{E}[B_C]$ as we have $0\leq\gamma\leq1$. It is also trivial to verify that $\rho_A+\rho_{pool}+\rho_{rest}=1$.

Let $B_O$ denote the total number of blocks found in an attack cycle. The block redundancy ratio of T-PAW is
\begin{align}
    \delta^{p_1,p_2,T}&=\frac{\mathbb{E}[B_O]}{\mathbb{E}[B_C]}\label{eq::withhold_cycle_dur_type_b},
\end{align}
where
\begin{align}
\mathbb{E}[B_O]=&\mathbb{E}[B_C]+\alpha p_1\Bigg(e^{-\lambda_2 T}\alpha p_2 \lambda_1 T\nonumber\\&+\big(1-e^{-\lambda_2 T}\big)\bigg(1+\alpha p_2 \lambda_1 T_e\bigg)\Bigg),
\end{align}
with
\begin{align}
 T_e&=\frac{1}{\lambda_2}-\frac{ T e^{-\lambda_2 T}}{1-e^{-\lambda_2 T}}.
\end{align}

The goal of the adversary is to optimize an objective function, such as the revenue ratio $\rho^{p_1,p_2,T}_{A}$ with respect to $p_1,p_2,T$. Here, an objective function such as $\rho^{p_1,p_2,T}_{A}$ is not concave or convex in general. In Appendix~\ref{sec::app::t-paw}, we provide proofs for all the quantities stated here and in Appendix~\ref{sec::app::paw-reduction}, for the sake of completeness, we take $T\to\infty$ in order to recover the quantities of PAW modeling of \cite{doger2025incentiveattacksbtcshortterm}. Similarly, we derive the T-PAW quantities using the PAW modeling of \cite{power_adjusting} with the simplified $c$ parameter in Appendix~\ref{sec::app::paw_simple}. Further, Appendix~\ref{sec::app::paw_simple} shows how to recover PAW, FAW and BWH attacks from T-PAW by setting specific $c,p_1,p_2$ and $T$ values. For the remainder of the paper, we simply assume that the pool manager is rational, i.e., $\mathds{1}_{P_R}=1$. In fact, if the pool manager is not rational, the improvements of T-PAW over PAW are even more stronger than what is presented in the rest of the paper.

\begin{remark}
It is trivial to show that the specific values of $\lambda_1$ and $T$ do not matter since expectations of the quantities $B_A,B_P,B_R,B_C,B_O$ depend on the product $\lambda_1 T$. Intuitively, this observation makes sense since the optimal withholding time $T$ should scale with the block production rate. Note that at $t_0$, $\lambda_1=\frac{D_0}{\tau_0}$ and the DAA changes the difficulty later at $t_1$ which in turn results in $\lambda_1=\frac{D_0}{\tau_0}\delta^{p_1,p_2,T}$ (and $\lambda_2$ scaled similarly). But we do not care about the change as we simply scale $T$ whenever $\lambda_1$ is scaled to keep $\lambda_1 T$ at the same level which in turn allows us to avoid re-solving the same problem of finding optimal $T$.
\end{remark}

\section{Numerical Results}
\begin{figure}[t]
    \centering
\begin{subfigure}[t]{0.48\columnwidth}
\centering
\includegraphics[width=\textwidth]{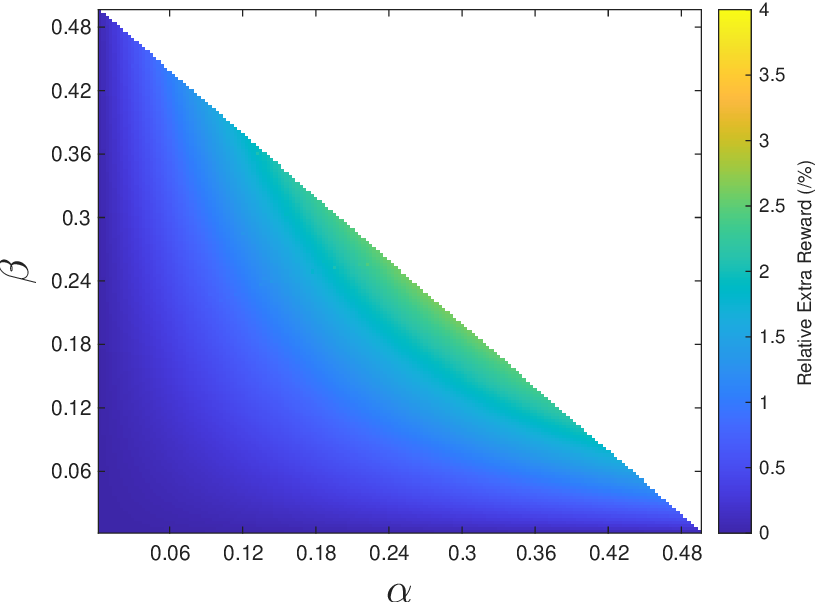}
    \caption{(PAW) $RER_A^{\dagger,H}$}
    \label{fig::RER_p_gamma_0}
\end{subfigure}
~
\begin{subfigure}[t]{0.48\columnwidth}
    \centering
    \includegraphics[width=\textwidth]{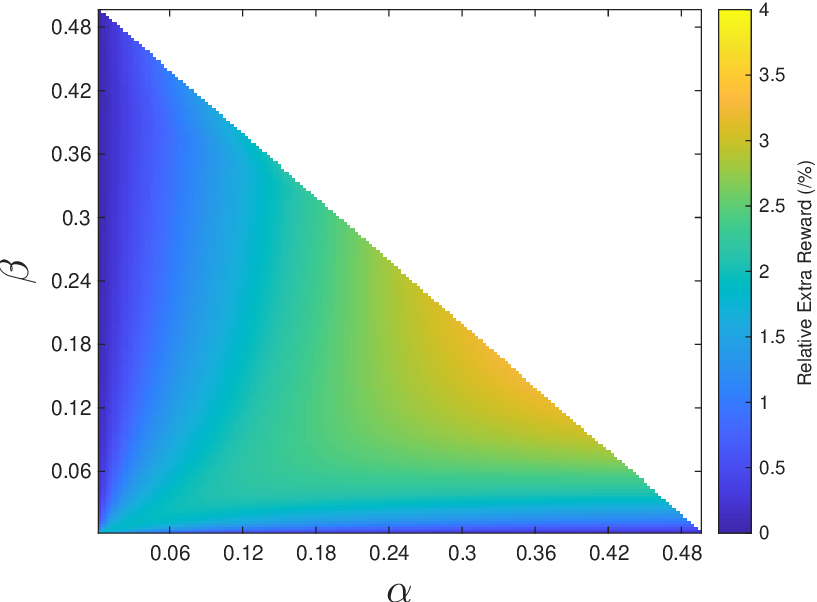}
    \caption{(T-PAW) $RER_A^{\ddagger,H}$}
    \label{fig::RER_t_gamma_0}
\end{subfigure}
~
\begin{subfigure}[t]{0.48\columnwidth}
\centering
\includegraphics[width=\textwidth]{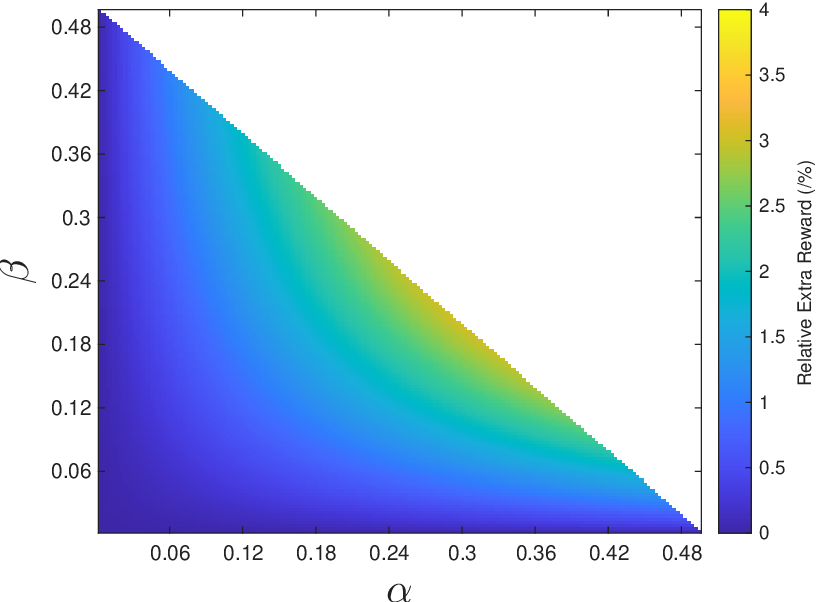}
    \caption{(PAW) $RER_{rest}^{\dagger,H}$}
    \label{fig::rest_RER_p_gamma_0}
\end{subfigure}
~
\begin{subfigure}[t]{0.48\columnwidth}
    \centering
    \includegraphics[width=\textwidth]{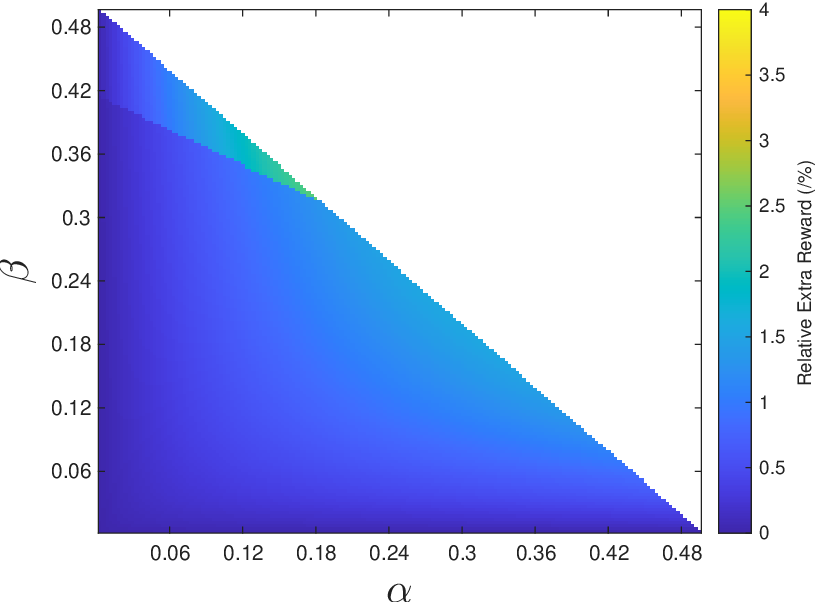}
    \caption{(T-PAW) $RER_{rest}^{\ddagger,H}$}
    \label{fig::rest_RER_t_gamma_0}
\end{subfigure}
~
\begin{subfigure}[t]{0.48\columnwidth}
    \centering
    \includegraphics[width=\textwidth]{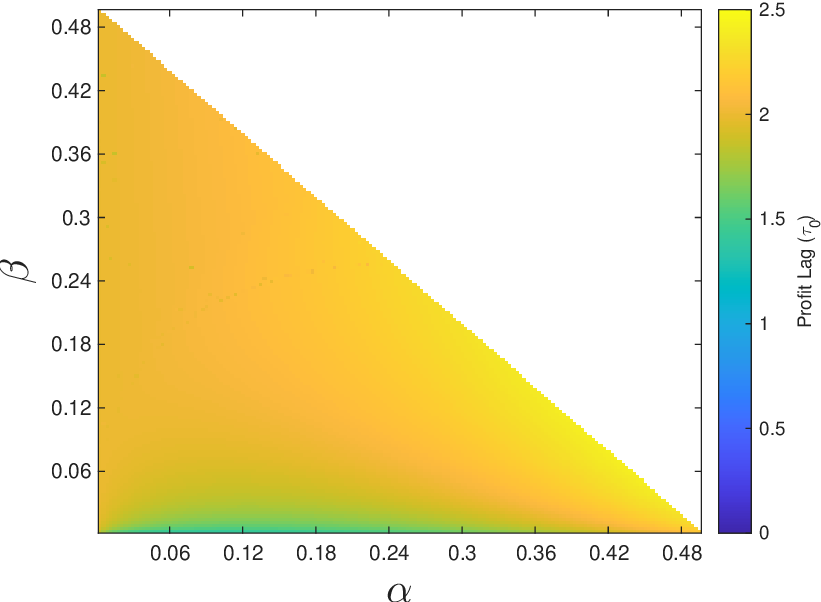}
    \caption{(PAW) $\Delta_{A}^{\dagger,H}$}
    \label{fig::lag_p_gamma_0}
\end{subfigure}
~
\begin{subfigure}[t]{0.48\columnwidth}
    \centering
    \includegraphics[width=\textwidth]{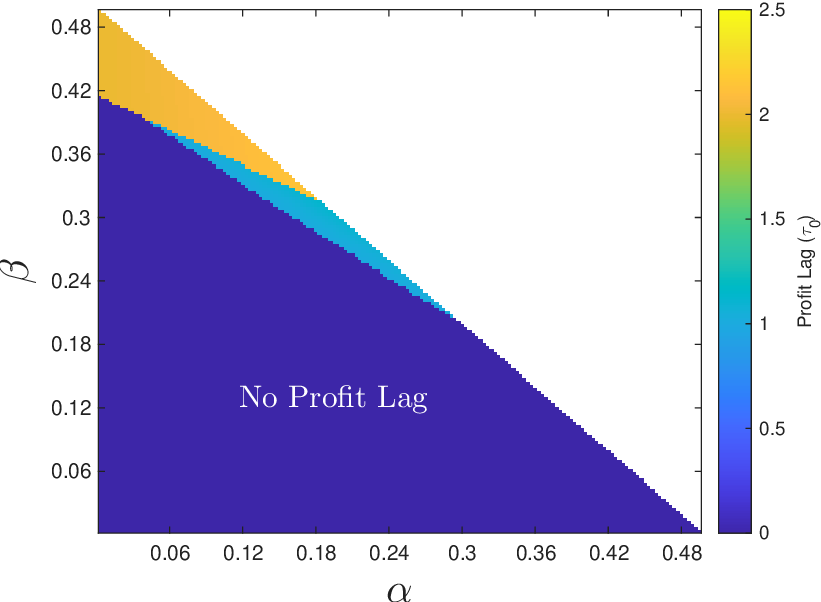}
    \caption{(T-PAW) $\Delta_{A}^{\ddagger,H}$}
    \label{fig::lag_t_gamma_0}
\end{subfigure}
~
\begin{subfigure}[t]{0.48\columnwidth}
    \centering
    \includegraphics[width=\textwidth]{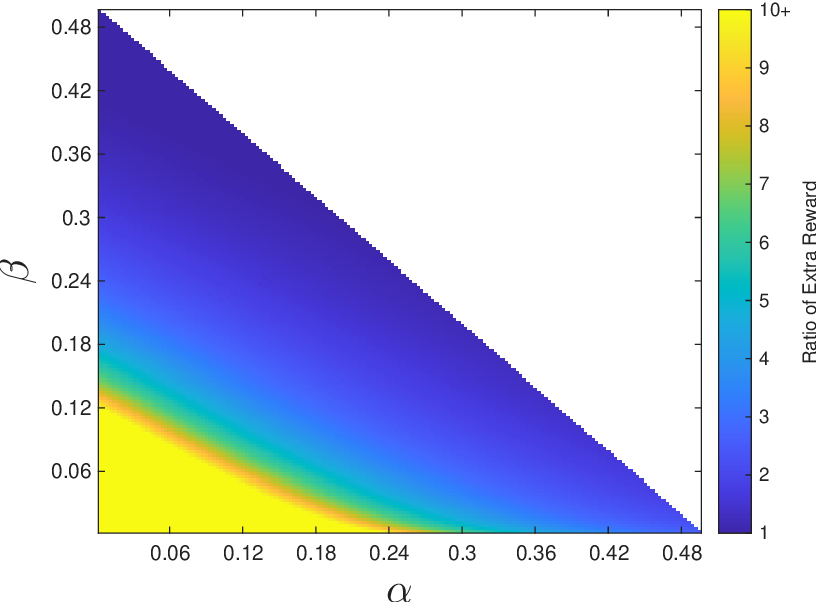}
    \caption{$\frac{RER_A^{\ddagger,H}}{RER_A^{\dagger,H}}$}
    \label{fig::RER_ratio_gamma_0}
\end{subfigure}
~
\begin{subfigure}[t]{0.44\columnwidth}
    \centering
    \includegraphics[width=\textwidth]{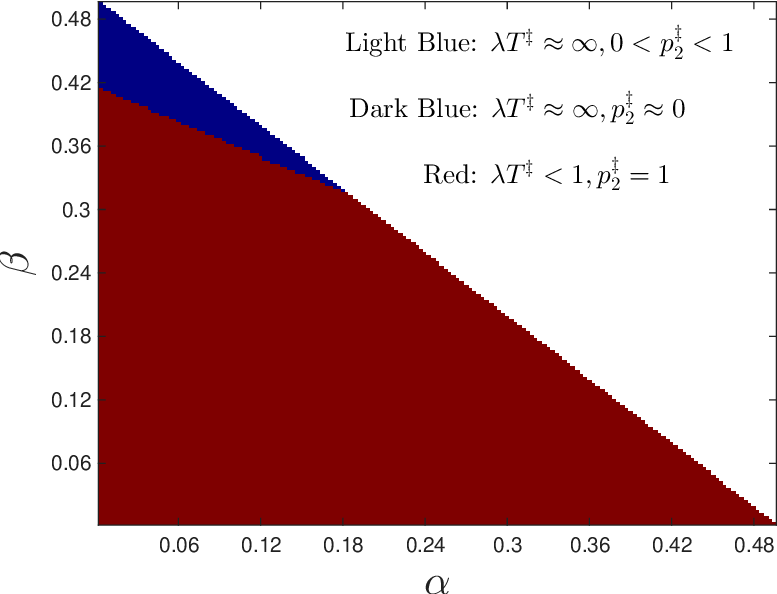}
    \caption{(T-PAW) Optimized Values}
    \label{fig::opt_vals}
\end{subfigure}
    \caption{Revenue ratio ($\rho_A$) maximization, $\gamma=0$.}
	\label{fig::RER_gamma_0}
\end{figure}
\begin{figure}[t]
    \centering
\begin{subfigure}[t]{0.48\columnwidth}
\centering
\includegraphics[width=\textwidth]{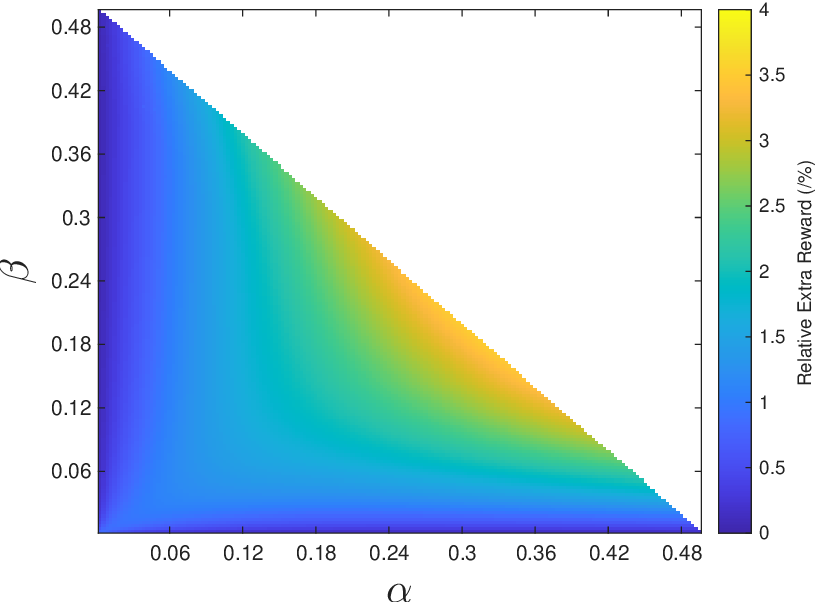}
    \caption{(PAW) $RER_A^{\dagger,H}$}
    \label{fig::RER_p_gamma_05}
\end{subfigure}
~
\begin{subfigure}[t]{0.48\columnwidth}
    \centering
    \includegraphics[width=\textwidth]{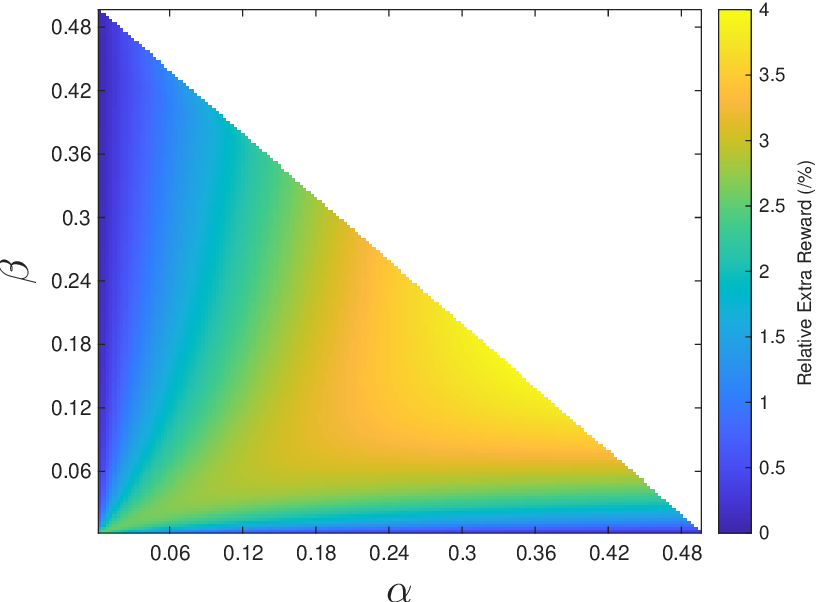}
    \caption{(T-PAW) $RER_A^{\ddagger,H}$}
    \label{fig::RER_t_gamma_05}
\end{subfigure}
~
\begin{subfigure}[t]{0.48\columnwidth}
\centering
\includegraphics[width=\textwidth]{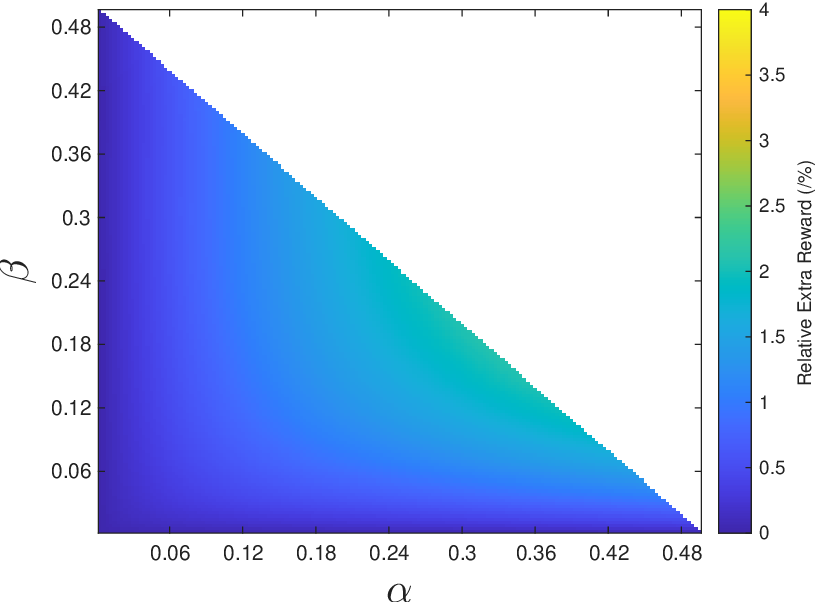}
    \caption{(PAW) $RER_{rest}^{\dagger,H}$}
    \label{fig::rest_RER_p_gamma_05}
\end{subfigure}
~
\begin{subfigure}[t]{0.48\columnwidth}
    \centering
    \includegraphics[width=\textwidth]{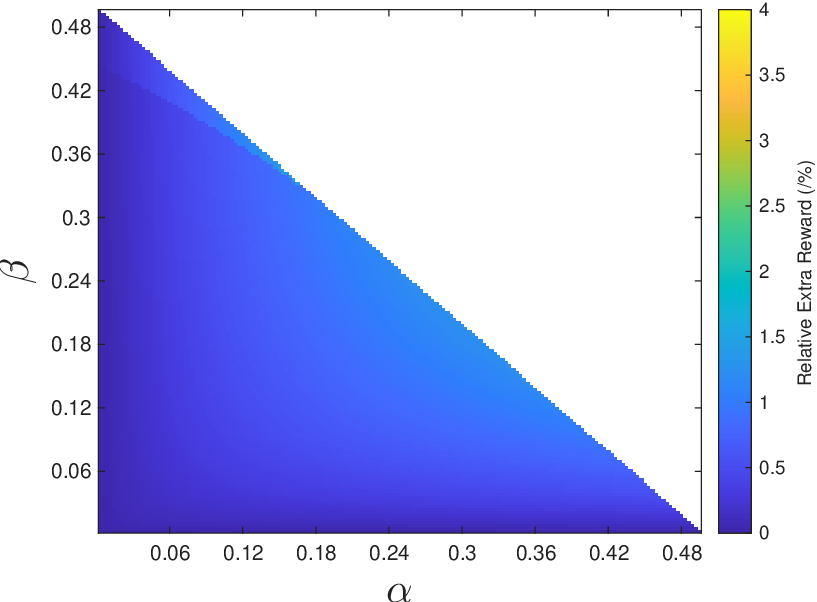}
    \caption{(T-PAW) $RER_{rest}^{\ddagger,H}$}
    \label{fig::rest_RER_t_gamma_05}
\end{subfigure}
~
\begin{subfigure}[t]{0.48\columnwidth}
    \centering
    \includegraphics[width=\textwidth]{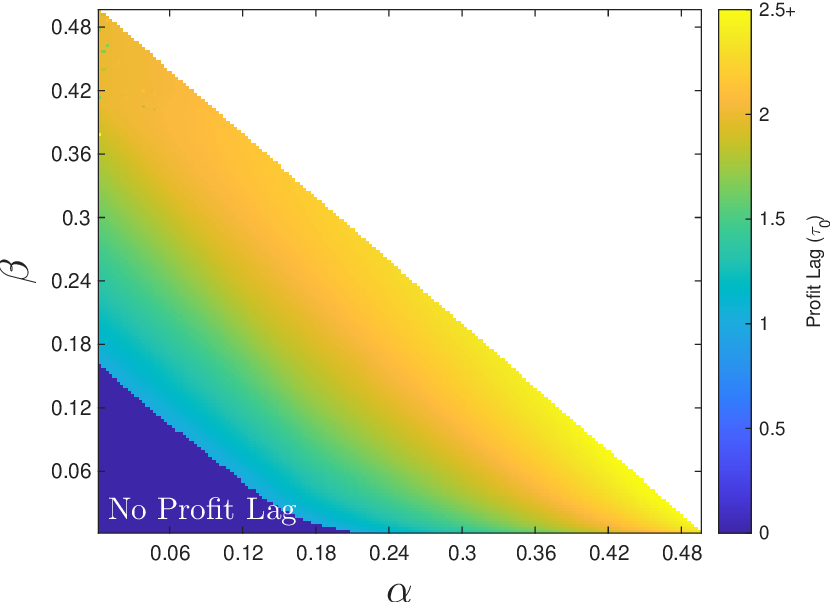}
    \caption{(PAW) $\Delta_{A}^{\dagger,H}$}
    \label{fig::lag_p_gamma_05}
\end{subfigure}
~
\begin{subfigure}[t]{0.48\columnwidth}
    \centering
    \includegraphics[width=\textwidth]{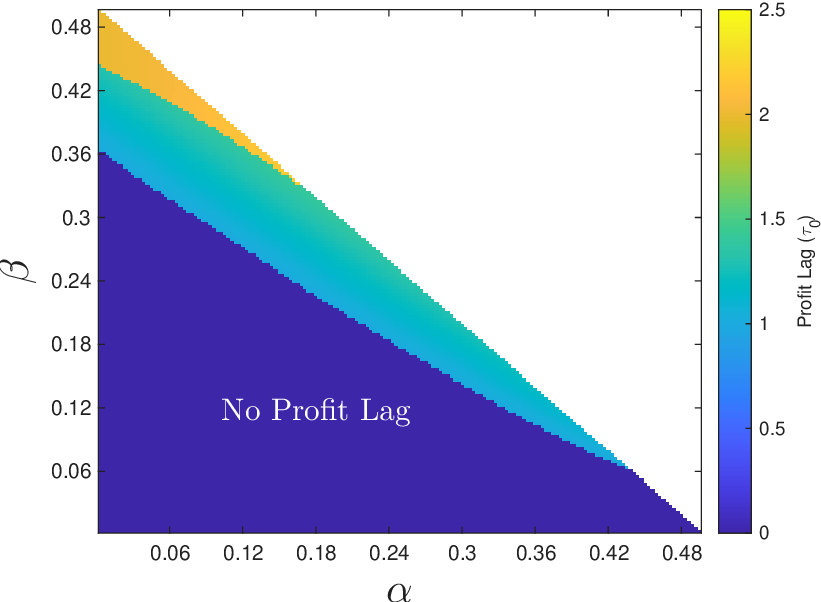}
    \caption{(T-PAW) $\Delta_{A}^{\ddagger,H}$}
    \label{fig::lag_t_gamma_05}
\end{subfigure}
~
\begin{subfigure}[t]{0.48\columnwidth}
    \centering
    \includegraphics[width=\textwidth]{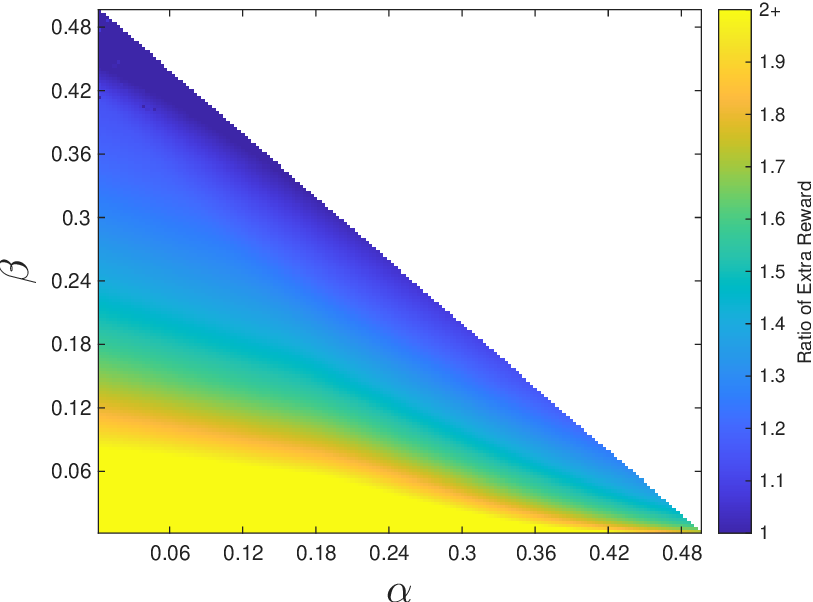}
    \caption{$\frac{RER_A^{\ddagger,H}}{RER_A^{\dagger,H}}$}
    \label{fig::RER_ratio_gamma_05}
\end{subfigure}
~
\begin{subfigure}[t]{0.44\columnwidth}
    \centering
    \includegraphics[width=\textwidth]{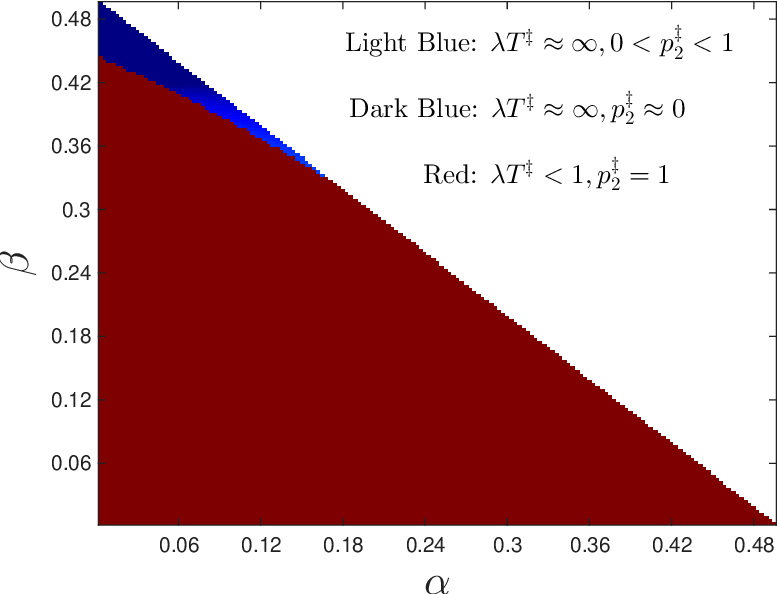}
    \caption{(T-PAW) Optimized Values}
    \label{fig::opt_vals_gamma_05}
\end{subfigure}
    \caption{Revenue ratio ($\rho_A$) maximization, $\gamma=0.5$.}
	\label{fig::RER_gamma_05}
\end{figure}

\begin{figure}[t]
     \centering
    \includegraphics[width=0.99\columnwidth]{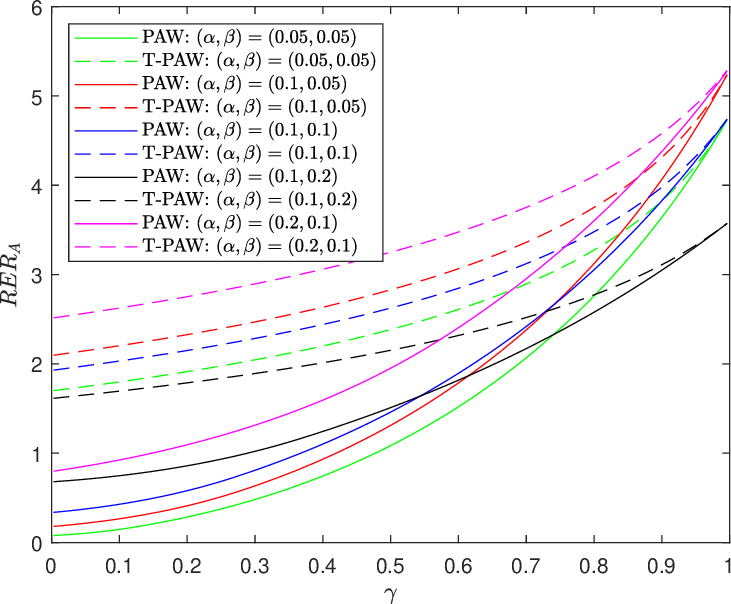}
    \caption{RER for PAW and T-PAW for 5 cases.}
    \label{fig::RER_gamma_line}
\end{figure}

We let $\dagger$ denote the PAW strategy ($p_1^{\dagger},p_2^{\dagger}$), and $\ddagger$ denote the T-PAW strategy ($p_1^{\ddagger},p_2^{\ddagger},T^{\ddagger}$) under the adversarial revenue ratio maximization, whereas $H$ denotes that the adversary uses the honest strategy ($T=0$). Similar to \cite{doger2025incentiveattacksbtcshortterm}, we also consider the maximization of $\Delta_A^{\star,H}(t_1)$ where the adversary adopts the T-PAW strategy $\star$ with $p^{\star}_1,p^{\star}_2,T^{\star}$ that maximizes the revenue change at the end of the first epoch (similarly PAW strategy $*$ denotes the maximization of revenue change at $t_1$ in PAW). For a given set of $\alpha,\beta,\gamma$ values, if $\exists (p_1,p_2,T)$ s.t. $\Delta_A^{\star,H}(t_1)>0$, this implies that the honest strategy is suboptimal compared to T-PAW.

\begin{remark}
    Since objective functions are not convex in general, we use optimization tools on MATLAB to first find local maximums of the objective function for a given set of $\alpha,\beta,\gamma$ values and take the maximum across local maxima obtained by $\mathsf{fmincon}$ starting from at least $100$ different initial values of $p_1,p_2,T$.
\end{remark}

\subsection{Numerical Maximization of $\rho_A$}
We pick $\gamma=\{0,0.5\}$ and present the RERs for the adversary and the honest non-pool miners for all values of $\alpha+\beta<0.5$ (with $\epsilon=0.003$ increments in $\alpha$ and $\beta$) for both PAW and T-PAW in \figref{fig::RER_gamma_0} and \figref{fig::RER_gamma_05} under the adversarial revenue ratio maximization. Comparing \figref{fig::RER_p_gamma_0} with \figref{fig::RER_t_gamma_0} (similarly \figref{fig::RER_p_gamma_05} vs \figref{fig::RER_t_gamma_05}), we clearly see the improvement of T-PAW over PAW. Notice, the RER of non-pool miners drop significantly from PAW (\figref{fig::rest_RER_p_gamma_0}) to T-PAW (\figref{fig::rest_RER_t_gamma_0}) when $\gamma=0$, which verifies our intuition that withholding too long in PAW risks the adversarial fPoW contributions (especially when $\gamma$ is small) as the fork race results more likely in the favor of non-pool miners. 

To directly compare the adversarial improvements of T-PAW over PAW, ratio of RERs are shown in \figref{fig::RER_ratio_gamma_0} and \figref{fig::RER_ratio_gamma_05}. Note that T-PAW improvements are especially strong when $\alpha,\beta,\gamma$ values are low as we proposed in Section~\ref{sec::discussion_tpaw_paw}. In fact, even though we only display the ratio values up to $10$ in \figref{fig::RER_ratio_gamma_0} (up to $2$ in \figref{fig::RER_ratio_gamma_05}), the extra reward of T-PAW grows further as $\alpha$ and $\beta$ values drop (e.g., when $\gamma=0$, the relative extra reward of T-PAW is 15 times the relative extra reward of PAW  when an adversary of size $\alpha=0.06$ attacks a pool of size $\beta=0.06$).  

In our analysis, we proved $\rho_{rest}\geq 1-\alpha-\beta$, i.e., non-pool miners benefit from the T-PAW attack. \cite{doger2025incentiveattacksbtcshortterm} shows that non-pool miners benefit from the PAW attack even more than the adversary that mounts the attack when $\gamma$ is low as seen when one compares \figref{fig::RER_p_gamma_0} with \figref{fig::rest_RER_p_gamma_0}. On the other hand, with our improvements, in T-PAW, non-pool miners benefit less than the adversary that mounts the attack even when $\gamma=0$ as seen when one compares \figref{fig::RER_t_gamma_0} with \figref{fig::rest_RER_t_gamma_0}. Moreover, \figref{fig::lag_p_gamma_0} vs \figref{fig::lag_t_gamma_0} (and \figref{fig::lag_p_gamma_05} vs \figref{fig::lag_t_gamma_05}) also display the adversarial profit lags in terms of $\tau_0$ for PAW vs T-PAW strategies, which hints that maximizing the adversarial revenue ratio in T-PAW results in no profit lag for a wide range of parameters. In other words, maximizing $\rho_A$ with the T-PAW strategy is superior to honest strategy for those parameters even when there is no DAA.

The RERs and profit lag curves suggest that for high $\beta$ values the optimal values of PAW and T-PAW overlap. In fact, when we investigate the optimal $p^{\dagger}_1,p^{\dagger}_2$ values obtained from PAW and $p^{\ddagger}_1,p^{\ddagger}_2,T^{\ddagger}$ from T-PAW, we notice that there is a diagonal boundary line as shown in \figref{fig::opt_vals} and \figref{fig::opt_vals_gamma_05} above which $\lambda_1 T^{\ddagger}\approx\infty$ and $p^{\dagger}_1=p^{\ddagger}_1,p^{\dagger}_2=p^{\ddagger}_2$. In other words, in these (non-red) regions, optimal T-PAW reduces to PAW. On the other hand, below the boundary (red region), we observe that $\lambda_1 T^{\ddagger}<1$ and $p^{\ddagger}_2=1$, i.e., the adversary decides for a temporary withholding and fully invests in increasing its shares for the fPoW since it knows that the fPoW will be shared soon. Further, the value of $\lambda_1 T^{\ddagger}$ inside the red region is monotonically increasing as $\beta$ increases, which verifies our initial intuition that the withholding saturates at some point (since, as pool size increases, the adversary needs more time until the fractional reward saturates). 

For the sake of completeness, in \figref{fig::RER_gamma_line}, for 5 cases of $(\alpha,\beta)$ values, we display RER of the adversary under PAW and T-PAW maximization when $\gamma$ is varying between $[0,1]$. Notice, the green curves, where $(\alpha,\beta)$ are small and when $\gamma$ is low, T-PAW strategy brings more than 20-fold improvement in RER. In fact, as we drop $(\alpha,\beta)$ RER of PAW diminishes completely whereas T-PAW always brings extra revenues. We note that, $\forall\gamma$ optimization results in $p^{\ddagger}_2=1$ in all 5 cases, whereas $p^{\dagger}_2=1$ only for high values of $\gamma$, i.e., when the risk of losing the fork race is low. On the other hand, $\lambda_1 T^{\ddagger}$ is monotonically increasing with $\gamma$, i.e., the adversary withholds more when the risk of losing the fork race is low and with $\gamma\to1$, T-PAW reduces to PAW. All these observations confirm our initial intuition about potential improvements of temporary withholding instead of $T\to\infty$. Another interesting observation in these $5$ cases is that, when $\gamma\to1$, the final maximized RER depends on $\frac{\alpha}{\beta}$.

\subsection{Revenue Change Maximization at $t_1$}
\begin{figure}[t]
    \centering
\begin{subfigure}[t]{0.48\columnwidth}
\centering
\includegraphics[width=\textwidth]{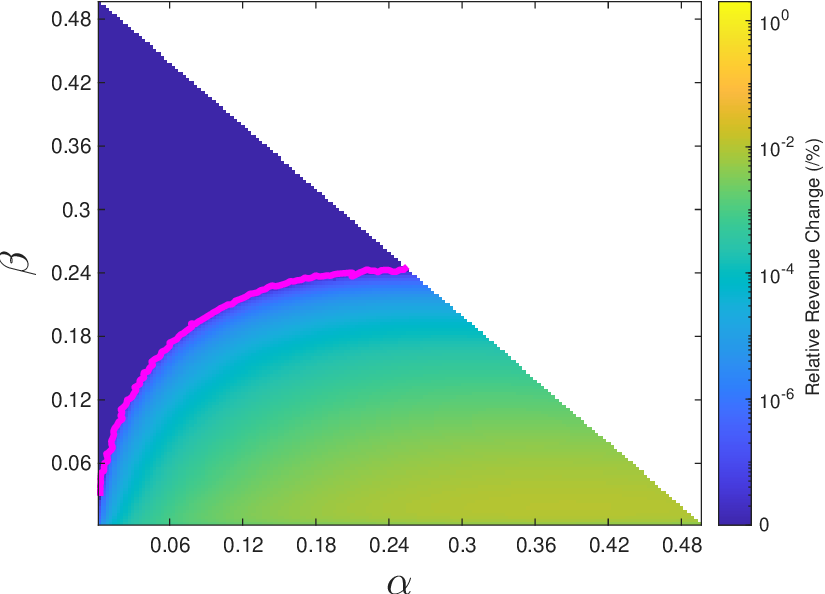}
    \caption{PAW, no profit lag below magenta, but less than $0.01\%$ gain}
    \label{fig::rel_rev_chan_at_t1_p}
\end{subfigure}
~
\begin{subfigure}[t]{0.48\columnwidth}
    \centering
    \includegraphics[width=\textwidth]{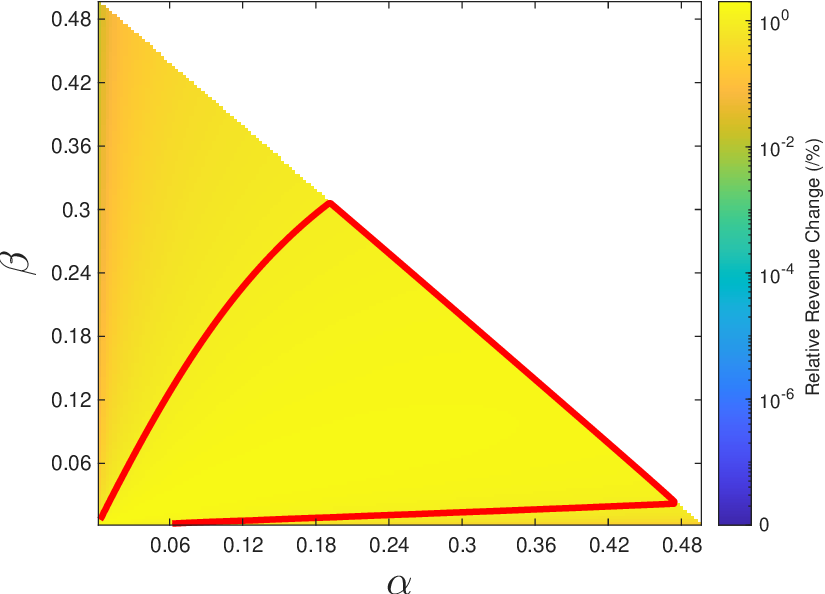}
    \caption{T-PAW, always no profit lag, more than $1\%$ gain inside red}
    \label{fig::rel_rev_chan_at_t1_t}
\end{subfigure}
    \caption{Revenue change maximization at $t_1$, $\gamma=0$.}
	\label{fig::rel_rev_chan_at_t1}
\end{figure}

Next, we consider the maximization of the revenue change at $t_1$ for PAW ($\overline\Delta_A^{*,H}(t_1)$) and T-PAW ($\overline\Delta_A^{\star,H}(t_1)$) and display the maximized relative values in \figref{fig::rel_rev_chan_at_t1} for $\gamma=0$. Magenta line in \figref{fig::rel_rev_chan_at_t1_p} is the boundary below which the maximization results in a (albeit small) positive revenue change, i.e., no profit lag, whereas above the boundary no $p_1,p_2$ values result in a positive revenue change before the difficulty adjustment. On the other hand, with T-PAW even when $\gamma=0$, the adversary always finds $p^{\star}_1,p^{\star}_2,T^{\star}$ values that result in positive revenue change before the difficulty adjustment and the gain is nontrivial, i.e., more than $0.1\%$. Further, inside the red marked region in \figref{fig::rel_rev_chan_at_t1_t}, the gain is above $1\%$.
\section{Discussion}
\subsection{Related Works and Future Directions}
Block withholding attacks have been initially investigated by Rosenfeld in \cite{rosenfeld2011analysisbitcoinpooledmining}, where the author specifically focuses on two versions, lie-in-wait and sabotage attacks. Sabotage attack is the case where the adversary withholds the fPoW to harm the pool revenues as in the case of the attack against Eligius mining pool \cite{EligiusBlockWithholding2014}. Later Courtois and Bahack show that the adversary can not only harm the pool revenues but also increase its own share of revenues in the system \cite{courtois2014subversiveminerstrategiesblock} with this attack (now known as BWH). FAW \cite{fork_after_witholding_attack} and PAW \cite{power_adjusting} attacks are sophistications of BWH as we explained earlier. On the other hand, lie-in-wait attack where the adversary temporarily withholds the fPoW and focuses on increasing its share (which can be seen as increasing $p_1$ to $p_2$) is dismissed by Courtois and Bahack as being not very realistic \cite{courtois2014subversiveminerstrategiesblock}. Contrary to the arguments of \cite{courtois2014subversiveminerstrategiesblock}, our analysis and generalization enables to bring all these attacks (BWH, FAW, PAW and lie-in-wait) together within the same framework and show that there is a tremendous improvement in doing so. 

Note that, a related study \cite{block_withholding_delay} investigates an attack with a deadline on fPoW submission time which roughly corresponds to $p_1=p_2$ case in T-PAW, i.e., no power adjusting, which makes it suboptimal compared to T-PAW (even suboptimal to honest mining absent any DAA). Similarly, another attack, called Return-After-Withholding (RAW) \cite{return-after-withholding} is a subcase of PAW with $p_2=0$ and $p_1$ optimized, and is therefore contained within the PAW framework as a suboptimal PAW strategy hence omitted here. On the other hand, Share Withholding (SWH) attack studied in \cite{chang2020sharewithholdingattackblockchain} deals with the adversarial manipulation of pPoW submission times in FAW for time-decaying payout schemes however the claims therein do not account for power adjusting strategies.

There are various countermeasures against PAW attacks as mentioned in \cite{power_adjusting,sarker2019awrs,chang2019silent_timestamping}, which are either not or less effective against T-PAW. For example, the time-stamp field proposal of \cite{power_adjusting} is less effective against T-PAW since the adversary can increase its shares even within limited $T$-time. On the other hand, silent time-stamping of \cite{chang2019silent_timestamping} can be completely circumvented as the adversary knows how long it is going to withhold and can pick the increase in nonce steps (e.g., the adversary tries $x_{i}=N_Tx_{i-1}$ in each nonce trial before withholding phase but $x_{i}=x_{i-1}+1$ during the withholding) or uses sybil identities. Anti-withholding reward system \cite{sarker2019awrs} proposes to increase rewards for fPoW compared to pPoW which is effective against BWH, FAW and PAW when $\gamma$ is low but not against T-PAW since the adversary eventually releases fPoW before anyone else does in most cases. A more detailed analysis against the countermeasures is left to future works.

Various follow-up works on FAW/PAW delve deeper into the subject for improvements or more sophisticated analysis. For example, Uncle Block Attack (UBA)\cite{uncle_block_attack} accounts the extra rewards in FAW from stale blocks in protocols such as Ethereum (PoW), \cite{Wang2023EFAW} combines the attack with eclipse attacks for improvements, \cite{power_adjusting,if_you_cant_beat_pay_them,bm-paw_attack} suggest combining it with bribery, whereas \cite{Zhu2022_revisit_faw} shows that the attacks give even more revenues when one models the system with high propagation delays. Although distinct from our primary focus, these works address related aspects of the problem and could be jointly considered to extend the scope and applicability of our work.

Eyal \cite{miners_dilemma} investigates BWH between competing pools to show that pools face a dilemma similar to that of prisoner's dilemma. Similarly both \cite{fork_after_witholding_attack,power_adjusting} investigate the situation further and show that the improvements of FAW/PAW limits the parameter regimes where the dilemma holds and where the larger pool can avoid it. They also study the FAW/PAW attack against multiple pool cases. We believe that investigating the above phenomenon in our rigorous and exact modeling T-PAW is the most natural and promising future direction.

\appendices
\section{Derivations of T-PAW Quantities}\label{sec::app::t-paw}
In this section, whenever we refer to quantities such as rewards, we mean their \textit{expected values}. We start by considering how an attack cycle ends from the start of an attack cycle:
\begin{enumerate}
\item with probability (w.p.) $1-\alpha-\beta$, an honest miner outside the target pool creates a block.
\item w.p. $\alpha(1-p_1)$, individual adversarial mining power creates a block (the adversary gets 1 full block reward).
\item w.p. $\beta$, a target pool member creates a block (the adversary gets $r_1$ fraction of the reward).
\item w.p. $\alpha p_1$, the adversary finds a fPoW from pool mining and withholds it. During the withholding phase, where the adversary allocates $p_2$ fraction of its power to pool mining,\label{case::withhold_phase_start}
\begin{enumerate}
    \item w.p. $e^{-\lambda_2 T}$, either no block is created, or all blocks created are created by adversarial pool mining and withheld, which can be ignored. The adversary releases the withheld fPoW after $T$ time (it gets $r_s$ fraction of the block reward).\label{case::success_withhold}
    \item w.p. $1-e^{-\lambda_2 T}$, at least one block is created by an entity other than the adversarial pool mining. Under this condition, ignoring the blocks created by adversarial pool mining, the first block was created, \label{case::unsuccess_withhold}
    \begin{enumerate}
        \item w.p. $\frac{\alpha(1-p_2)}{1-\alpha p_2}$, by the individual mining power of the adversary (the adversary gets 1 block reward).
        \item w.p. $\frac{\beta}{1-\alpha p_2}$, a target pool member (the adversary gets $r_u$ fraction of the block reward).
        \item w.p. $\frac{1-\alpha-\beta}{1-\alpha p_2}$, by an honest miner outside the target pool. The adversary releases the withheld fPoW and a fork race starts. During the fork race phase,
        \begin{enumerate}
            \item w.p. $\alpha$, the adversary mines a new block on top of the recently released fPoW (the adversary gets 1 block reward and $r_u$ fraction of the fPoW reward).
            \item w.p. $\beta$, target pool mines a new block on top of the recently released fPoW (no adversarial contribution for this new block but the adversary gets $r_u$ fraction of the recently released fPoW).\footnote{Here, we assume a rational pool manager. If not, the pool and the honest miners outside the pool get 1 block reward each (no adversarial reward).}
            \item w.p. $1-\alpha-\beta$, honest miners outside the pool mine a new block (w.p. $\gamma$, this new block is mined on top of the recently released fPoW and the adversary gets $r_u$ fraction of the fPoW reward).
        \end{enumerate}
    \end{enumerate}
\end{enumerate}    
\end{enumerate}
The above cases fully cover an attack cycle and each cycle is independent. Thus, summing the product of the probabilities and the corresponding rewards of each entity, it is trivial to verify the quantities $\mathbb{E}[B_A]$ in \eqref{eq::adv_rho_bwh_type_b_t_Ba}, $\mathbb{E}[B_P]$ in \eqref{eq::adv_rho_bwh_type_b_t_Bp}, $\mathbb{E}[B_R]$ in \eqref{eq::adv_rho_bwh_type_b_t_Br} and $\mathbb{E}[B_C]$ in \eqref{eq::canonical_bwh_type_b_t} (see \figref{fig::cycle_events}). Next, we derive the quantities $r_1$, $r_s$ and $r_u$.

At the start of an attack cycle, let $\lambda_{(1,0)}=\alpha p_1\lambda_1$, $\lambda_{(1,2)}=\alpha(1-p_1)\lambda_1$, $\lambda_{(1,3)}=\beta\lambda_1$, $\lambda_{(1,4)}=(1-\alpha -\beta)\lambda_1$ denote the block arrival rates for the adversarial pool mining, adversarial individual mining, honest-pool mining and the honest-outside-pool mining and let $T_{(1,1)},T_{(1,2)},T_{(1,3)},T_{(1,4)}$ be the arrival time of the first block of the respective type with $T_1=\min(T_{(1,1)},T_{(1,2)},T_{(1,3)},T_{(1,4)})\sim\mathrm{Exp}(\lambda_1)$. If the target pool miners are the first to find a block in an attack cycle, the fraction of the reward the adversary gets from the fPoW is trivial:
\begin{align}
    r_1=\mathbb{E}\bigg[\frac{\alpha p_1T_{(1,3)}}{a_1T_{(1,3)}}\bigg|T_{(1,3)}=T_1\bigg]=a'_1.
\end{align}
On the other hand, with probability
\begin{align}
    \mathbb{P}(T_1=T_{(1,1)})=\alpha p_1,
\end{align}
the attack cycle enters a withholding phase. During the withholding phase (assume the withholding phase starts at time zero for the sake of simplicity), let $\lambda_{(2,1)}=\alpha p_2\lambda_1$, $\lambda_{(2,2)}=\alpha(1-p_2)\lambda_1$, $\lambda_{(2,3)}=\beta\lambda_1$ and $\lambda_{(2,4)}=(1-\alpha -\beta)\lambda_1$ denote the block arrival rates for the adversarial pool mining, adversarial individual mining, honest-pool mining and the honest-outside-pool mining, respectively. Similarly, let $T_{(2,1)},T_{(2,2)},T_{(2,3)},T_{(2,4)}$ be the arrival time of the first block of the respective type starting from the withholding phase and $T_2=\min(T_{(2,2)},T_{(2,3)},T_{(2,4)})$. It is clear that $T_{(1,\cdot)}$ are independent of $T_{(2,\cdot)}$ and $T_2\sim\mathrm{Exp}(\lambda_2)$, i.e.,
\begin{align}
    \mathbb{P}(T_2> T)= e^{-(1-\alpha p_2)\lambda_1 T}=e^{-\lambda_2 T}.
\end{align}
After withholding the fPoW for a time $T$ where no block was created, or all blocks that were created were created by adversarial pool mining, the adversary releases the fPoW. Since the adversary contributes with $\alpha p_1$ power to pool mining from the start of an attack cycle until the start of the withholding phase (lasts $T_{(1,1)}$ time), and contributes with $\alpha p_2$ power to pool mining in the withholding phase (lasts $T$ time), the fraction of the reward the adversary gets from the fPoW in Case~\ref{case::success_withhold} is
\begin{align}
     r_s&=\mathbb{E}\bigg[\frac{\alpha p_1T_{(1,1)}+\alpha p_2 T}{a_1T_{(1,1)}+a_2T}\bigg|T_{(1,1)}=T_1,T<T_2\bigg]\\
     &=\mathbb{E}\bigg[\frac{\alpha p_1T_1+\alpha p_2 T}{a_1T_1+a_2T}\bigg]\\
     &=a'_1\mathbb{E}\bigg[\frac{a_1T_1}{a_1T_1+a_2T}\bigg]+a'_2\mathbb{E}\bigg[\frac{a_2T}{a_1T_1+a_2T}\bigg]\\
     &=a_1'+(a_2'-a_1')\mathbb{E}\bigg[\frac{a_2T}{a_1T_1+a_2T}\bigg].
\end{align}
Note that $T'=a_2T$ is a deterministic scalar, $Q_1=a_1T_1\sim \mathrm{Exp}(\lambda'_1)$ and 
\begin{align}
    \mathbb{E}\bigg[\frac{a_2T}{a_1T_1+a_2T}\bigg]&=\mathbb{E}\!\left[\frac{T'}{Q_1 + T'}\right]\\ 
&= \int_0^\infty \frac{T'}{x+T'} \, \lambda'_1 e^{-\lambda'_1 x}\, dx\\
&=\lambda'_1 T' e^{\lambda'_1 T'}E_{1}(\lambda'_1 T').
\end{align}
On the other hand, during the withholding phase, with probability
\begin{align}
    \mathbb{P}(T_{2}< T)= 1-e^{-\lambda_2 T}
\end{align}
at least one block is created by an entity other than the adversarial pool mining. Let $T'_{2} \sim (T_2|T_2<T)$, in other words, $T'_{2} \sim \mathrm{Exp}(\lambda_2)\ \text{truncated to } [0,T]$. As a result, in the subcases of Case~\ref{case::unsuccess_withhold}, the fraction of the reward the adversary (if it ever) gets from the fPoW is
\begin{align}
     r_u&=\mathbb{E}\bigg[\frac{\alpha p_1T_{(1,1)}+\alpha p_2 T_2}{a_1T_{(1,1)}+a_2T_2}\bigg|T_{(1,1)}=T_1,T_2<T\bigg]\\
     &=\mathbb{E}\bigg[\frac{\alpha p_1T_1+\alpha p_2 T'_2}{a_1T_1+a_2T'_2}\bigg]\\
     &=a'_1\mathbb{E}\bigg[\frac{a_1T_1}{a_1T_1+a_2T'_2}\bigg]+a'_2\mathbb{E}\bigg[\frac{a_2T'_2}{a_1T_1+a_2T'_2}\bigg]\\
     &=a_1'+(a_2'-a_1')\mathbb{E}\bigg[\frac{a_2T'_2}{a_1T_1+a_2T'_2}\bigg].
\end{align}
Let $Q_1=a_1T_1\sim \mathrm{Exp}(\lambda'_1)$ and $Q_2=a_2T'_2\sim \mathrm{Exp}(\lambda'_2)\ \text{truncated to } [0,T']$, i.e., the density of $Q_2$ is
\begin{align}
f_{Q_2}(y) = \frac{\lambda'_2 e^{-\lambda'_2 y}}{1 - e^{-\lambda'_2 T'}}, 
\quad 0 \le y \le T'.
\end{align}
Then,
\begin{align}
\mathbb{E}\bigg[\frac{a_2T'_2}{a_1T_1+a_2T'_2}\bigg]&=\mathbb{E}\!\left[\frac{Q_2}{Q_1 + Q_2}\right]\\&=\mathbb{E}\!\left[\mathbb{E}\!\left[\frac{t}{Q_1 + t}\bigg|Q_2=t\right]\right]\\
&=\mathbb{E}\!\left[ \lambda'_1 Q_2 \, e^{\lambda'_1 Q_2} E_1(\lambda'_1 Q_2)\right]\\
&= \frac{\lambda'_1 \lambda'_2}{1 - e^{-\lambda'_2 T'}} 
\int_0^{T'} t \, e^{(\lambda'_1 - \lambda'_2)t} E_1(\lambda'_1 t) dt.\label{eq::truncated_integral}
\end{align}

Finally, we derive the expected quantity of the orphan blocks in an attack cycle, $B_O-B_C$, i.e., blocks that are mined in an attack cycle but do not become a canonical block. Let $T_e=\mathbb{E}[T'_{2}]$, i.e.,
\begin{align}
T_e&=\mathbb{E}[T'_{2}]=\frac{1}{\lambda_2}-\frac{Te^{-\lambda_2 T}}{1-e^{-\lambda_2 T}}.
\end{align}
Note that block withholding only happens in subcases of Case~\ref{case::withhold_phase_start} (w.p. $\alpha p_1$), which in turn results in orphan blocks. In Case~\ref{case::success_withhold}, the adversary will release one fPoW which becomes a canonical block however the other fPoWs created by the adversarial pool mining will be discarded. Hence, w.p. $e^{-\lambda_2 T}$, $N_{4a}$ blocks do not become a canonical blocks in this case, where $N_{4a}\sim \text{Poi}(\lambda_{(2,1)} T)$. On the other hand, in Case~\ref{case::unsuccess_withhold}, a fPoW mined by the adversary is forked by another entity's block (or by adversarial individual mining), hence, at least one block becomes orphan in addition to the other fPoWs created by the adversarial pool mining that are discarded in the withholding phase. Note that the discarding happens for $T'_{2}$ duration time, i.e., until the arrival time of the first block during the withholding phase. Hence, w.p. $1-e^{-\lambda_2 T}$, $N_{4b}$ additional fPoWs do not become a canonical blocks in this case where $N_{4b}\sim \text{Poi}(\lambda_{(2,1)} T'_{2})$. Thus,
\begin{align}
    \mathbb{E}[B_O]=&\mathbb{E}[B_C]+\alpha p_1\Bigg(e^{-\lambda_2 T}\mathbb{E}[N_{4a}]\nonumber\\&+\big(1-e^{-\lambda_2 T}\big)\Big(1+\mathbb{E}[N_{4b}]\Big)\Bigg),
\end{align}
which completes the proof since $\mathbb{E}[N_{4a}]= \alpha p_2 \lambda_1 T$ and $\mathbb{E}[N_{4b}]= \alpha p_2 \lambda_1 T_e$.

\section{Reduction to PAW When $T\to\infty$}\label{sec::app::paw-reduction}
Notice, when $T\to\infty$, Case~\ref{case::success_withhold} never happens. Thus, we are only interested in subcases of Case~\ref{case::unsuccess_withhold} where $T\to\infty$ which implies $T'_{2}\xrightarrow{\mathrm{}} T_2\sim\mathrm{Exp}(\lambda_2)$. Let us denote the average fraction of the reward the adversary gets from the fPoW as $r_\infty$, i.e.,
\begin{align}
    \lim_{T\to\infty}r_u=r_\infty,
\end{align}
which we will derive later explicitly.
Note,
\begin{align}
    \rho_A^{p_1,p_2}&=\lim_{T\to\infty}\rho^{p_1,p_2,T}_{A}\\
    &=\frac{\lim_{T\to\infty}\mathbb{E}[B_A]}{\lim_{T\to\infty}\mathbb{E}[B_C]}\label{eq::adv_rho_bwh_type_b_1}\\
    &=\frac{\mathbb{E}[\lim_{T\to\infty}B_A]}{\mathbb{E}[\lim_{T\to\infty}B_C]},\label{eq::adv_rho_bwh_type_b_2}
\end{align}
where \eqref{eq::adv_rho_bwh_type_b_1} follows from the law of large numbers and \eqref{eq::adv_rho_bwh_type_b_2} follows from the dominated convergence theorem since $\lvert B_A\rvert<\lvert B_C\rvert \leq 2$. Hence,
\begin{align}
\mathbb{E}&[\lim_{T\to\infty}B_A] =  \alpha(1-p_1) + \beta r_1 + \alpha p_1\Big( \frac{\alpha(1 - p_2)}{1 - \alpha p_2} +  \frac{\beta}{1 - \alpha p_2}r_\infty  \nonumber \\
&+\frac{1-\alpha-\beta}{1 - \alpha p_2}\big(r_\infty(\gamma(1-\alpha-\beta)+\alpha+\beta\mathds{1}_{P_R})+\alpha\big)\Big),\\
\mathbb{E}&[\lim_{T\to\infty}B_C]=1+\alpha p_1\frac{1-\alpha-\beta}{1 - \alpha p_2}.
\end{align}
Similar arguments hold for other quantities with
\begin{align}
\mathbb{E}&[\lim_{T\to\infty}B_P] =   \beta (1-r_1) + \alpha p_1 \Big( \frac{\beta}{1 - \alpha p_2}(1-r_\infty)+\nonumber \\ 
&\frac{1-\alpha-\beta}{1 - \alpha p_2}\big((1-r_\infty)(\gamma(1-\alpha-\beta)+\alpha+\beta\mathds{1}_{P_R})+\beta\big)\Big),\\
\mathbb{E}&[\lim_{T\to\infty}B_R] = (1-\alpha-\beta)\nonumber\\
&\times\bigg(1+\frac{\alpha p_1}{1-\alpha p_2}\big((1-\alpha-\beta)(2-\gamma)+\beta \overline{\mathds{1}_{P_R}})\bigg),\\
\mathbb{E}&[\lim_{T\to\infty}B_O]=\mathbb{E}[\lim_{T\to\infty}B_C]+\frac{\alpha p_1}{1-\alpha p_2}.
\end{align}
We recovered the quantities of PAW given in s\cite{doger2025incentiveattacksbtcshortterm} by taking $T\to\infty$ for the quantities given in Section~\ref{sec::t-paw} except that we have not given an explicit formula for $r_\infty$, which is done next as
\begin{align}
    r_\infty&=\lim_{T\to\infty}\mathbb{E}\bigg[\frac{\alpha p_1T_{(1,1)}+\alpha p_2 T_2}{a_1T_{(1,1)}+a_2T_2}\bigg|T_{(1,1)}=T_1,T_2<T\bigg]\\
    &=\mathbb{E}\bigg[\frac{\alpha p_1T_1+\alpha p_2 T_2}{a_1T_1+a_2T_2}\bigg]\\
    &=a'_1\mathbb{E}\bigg[\frac{a_1T_1}{a_1T_1+a_2T_2}\bigg]+a'_2\mathbb{E}\bigg[\frac{a_2T_2}{a_1T_1+a_2T_2}\bigg]\\
    &=a_1'+(a_2'-a_1') \mathbb{E}\bigg[\frac{a_2T_2}{a_1T_1+a_2T_2}\bigg],
\end{align}
where $a_1T_1\sim \mathrm{Exp}(\lambda'_1)$, $a_2T_2\sim \mathrm{Exp}(\lambda'_2)$ and we leave as an exercise to verify (using $T'=\infty$ in \eqref{eq::truncated_integral}), when $\lambda'_1\neq \lambda'_2$,
\begin{align}
    \mathbb{E}\bigg[\frac{a_2T_2}{a_1T_1+a_2T_2}\bigg]=\frac{\lambda'_1}{(\lambda'_1-\lambda'_2)^2}\bigg(\lambda'_2 \ln\frac{\lambda'_2}{\lambda'_1}-\lambda'_2+\lambda'_1 \bigg),
\end{align}
and $1/2$ by symmetry when $\lambda'_1= \lambda'_2$. Notice, with $T\to\infty$, the authors of \cite{power_adjusting} and \cite{doger2025incentiveattacksbtcshortterm} make an approximation as follows
\begin{align}
    \mathbb{E}\bigg[\frac{\alpha p_1T_1+\alpha p_2 T_2}{a_1T_1+a_2T_2}\bigg]&\approx\frac{\alpha p_1 \mathbb{E}[T_1]+\alpha p_2 \mathbb{E}[T_2]}{a_1 \mathbb{E}[T_1]+a_2 \mathbb{E}[T_2]}\\
    &=\frac{\alpha p_1/\lambda_1 +\alpha p_2/\lambda_2}{a_1 /\lambda_1+a_2 /\lambda_2}\\
    &=\frac{\alpha \overline p}{\beta+\alpha \overline p},
\end{align}
where $\overline p=\frac{p_1+p_2-\alpha p_1p_2}{2-\alpha p_2}$ when deriving $r_{\infty}$ which is only an approximation and the error grows large especially if $T\neq \infty$.

\section{Simplified T-PAW Analysis}\label{sec::app::paw_simple}
Assume each fork race ends in the favor of the adversary with probability $c$ and each attack cycle is independent as in \cite{power_adjusting,fork_after_witholding_attack}. Then, for the adversarial revenue ratio, we get
\begin{align}
    \rho^{p_1,p_2,T}_{A}=&\alpha(1-p_1)+\beta r_1+\alpha p_1\bigg(e^{-\lambda_2 T}r_s +\big(1-e^{-\lambda_2 T}\big) \nonumber \\ &\times\bigg(\frac{\alpha(1-p_2)}{1-\alpha p_2}+r_u\frac{\beta+c(1-\alpha-\beta)}{1-\alpha p_2}\bigg)\bigg),
\end{align}
for the honest pool members revenue ratio, we get
\begin{align}
    \rho_{pool}=&\beta(1-r_1)+\alpha p_1\bigg(e^{-\lambda_2 T}(1-r_s)+\big(1-e^{-\lambda_2 T}\big) \nonumber \\ &\times\bigg((1-r_u)\frac{\beta+c(1-\alpha-\beta)}{1-\alpha p_2}\bigg)\bigg),
\end{align}
and for the rest of the honest miners, we get
\begin{align}
    \rho_{rest}=(1-\alpha-\beta)\left(1+\big(1-e^{-\lambda_2 T}\big)(1-c)\frac{\alpha p_1}{1-\alpha p_2}\right).
\end{align}
It is trivial to show that the results above reduce to those obtained for PAW in \cite{power_adjusting} when $T\to\infty$ (with a more rigorous derivation of $r_\infty$). On the other hand, setting $p_1=p_2$ with arbitrary $T$ can be called Temporary FAW attack whereas $p_1=p_2$ and $T\to\infty$ results in the same FAW analysis as in \cite{fork_after_witholding_attack} (note, when $p_1=p_2$, $r_s=r_u=r_1=a'_1$). Setting $p_1=p_2$ and $c=0$ with arbitrary $T$ can be called Temporary BWH attack whereas $p_1=p_2$, $c=0$ and $T\to\infty$ results in the same BWH analysis as in \cite{power_splitting_pools}.
\bibliographystyle{IEEEtran}
\bibliography{blockchain}
\end{document}